\documentclass[12pt,preprint]{aastex}

\begin{document}


\title{The Mass Dependence of Stellar Rotation in the Orion Nebula Cluster}


\author{W. Herbst}
\affil{Astronomy Department, Wesleyan University, Middletown, CT 06459}
\email{wherbst@wesleyan.edu}

\and

\author{C. A. L. Bailer-Jones and R. Mundt}
\affil{Max-Planck-Institut f\"ur Astronomie, K\"onigstuhl 17, 
D-69117 Heidelberg Germany}




\begin{abstract}
We have determined new rotation periods for 404 stars in the 
Orion Nebula Cluster using the Wide Field Imager attached to the 
MPG/ESO 2.2 m telescope on La Silla, Chile. Mass estimates are 
available for 335 of these and most have M $<$ 0.3 M$_{\odot}$.
We confirm the existence of a bimodal period distribution for the 
higher mass stars in our sample and show that the median rotation rate 
decreases with increasing mass for stars in the range 0.1$<$M$<$0.4 M$_{\odot}$. 
While the spread in angular momentum (J) at any given mass is more than a 
factor of 10, the majority of lower mass stars in the ONC rotate at rates 
approaching 30\% of their critical break-up velocity, as opposed to 5-10\% for 
solar-like stars. This is a consequence of {\it both} a small increase in 
observed specific angular momentum (j=J/M) and a larger decrease in the 
critical value 
of j with decreasing mass. Perhaps the most striking fact, however, is that 
j varies by so little - less than a factor of two - over the interval  
0.1-1.0 M$_{\odot}$. The distribution of rotation rates with 
mass in the ONC (age$\sim$1 My) is similar in nature to what is found in the 
Pleiades (age$\sim$100 My). These observations provide a significant 
new guide and test for models of stellar angular momentum evolution 
during the proto-stellar and pre-main sequence phases.
\end{abstract}




\section{Introduction}

Rotation is an undoubtedly important but often neglected aspect of stellar
evolution.  Our knowledge of the angular momentum evolution of stars 
and, in particular,
how it depends on mass is seriously incomplete, especially for low 
mass stars. This leads to uncertainty
about the importance of a variety of physical phenomena including internal
angular momentum transport, stellar winds and disk-locking in 
solar-like (M$\sim$0.5-1.5 M$_\odot$) and lower mass stars \citep{kri97}.  
Recent advances in theory (Sills, Pinsonneault and Terndrup 2000)
and observation \citep{ter00,bjm01} have allowed us
to investigate these issues to the H-burning limit and beyond.  Stars with 
M$<$0.5 M$_\odot$ are of particular interest because they are
fully convective and, therefore, likely to remain
rigid rotators for a Hubble time, once they achieve that status.  What has been
missing in these studies is a knowledge of the ``initial" 
(i.e.
at $\sim$1 My) distribution of angular momentum as a function of mass,
particularly for very low mass stars.  This is important as a starting
point for the studies mentioned and also as a test or guide for theories of
proto-star and early pre-main sequence (PMS) evolution.  

As usual, the best
way to observe the time and mass dependence of a stellar parameter is by
studying clusters.  Investigations of stellar rotation have relied on a few
nearby clusters, especially IC 2602, $\alpha$ Per, the Pleiades and the
Hyades, with ages ranging from 50 to 600 My.  Angular velocities are
directly determined by photometric monitoring.  Large cool spots produce
cyclical light variations as they rotate into and out of the line of sight. 
A Web-based compendium of photometric rotation periods for stars in these
clusters has been compiled by C. F. Prosser and J. R. 
Stauffer\footnote{available at 
http://cfa-www.harvard.edu/$\sim$stauffer/opencl/index.html}.  The Orion Nebula Cluster (ONC) represents an extremely
important addition to these data for the following reasons: 1) its age is
only $\sim$0.8 My \citep{hil97}, placing it among the youngest
clusters known, 2) it is sufficiently dense and well populated that it will
probably emerge from the formation process as a gravitationally bound
entity \citep{hh98}, making it directly comparable to the
other clusters used in rotation studies, and 3) it is well populated with very low mass
stars and brown dwarf candidates \citep{hil97,hc00}.

Rotational studies of low mass stars in the ONC were pioneered by 
\citet{mh91} and \citet{ah92} who first showed that the
solar-like stars had a bimodal distribution of rotation periods, similar to
what is found in this mass range for the 50-150 My old clusters.  Recent
work has extended our knowledge of rotation to a larger mass range within
the ONC and to the distributed population of pre-main sequence stars
located throughout the Orion A molecular cloud 
\citep{ch96,sta99,her00,reb01,ch01}.  Here we describe a further important extension to very
low mass stars, all of which are within the ONC defined on dynamical
grounds by \citet{hh98}.  This is the relevant portion of
the Orion association for direct comparison with older open clusters since
it is the only part that will likely maintain its identity for 50 My or
longer.  Also, it is important to isolate a sample with as small an 
age range as possible in
these studies, since rotation periods may evolve rapidly during
the first $\sim$1-2 My of a star's life.  For these reasons it is best to
focus on the ONC, as opposed to the entire Orion A population (or other T
associations), when studying the time dependence of stellar angular momentum
using clusters.

\section{Observations}

Ninety-two images of the ONC were obtained through an intermediate band
filter ($\lambda_c$=815.9 nm; $\Delta \lambda_{FWHM}$=20.9 nm; selected to
exclude strong nebular lines) on 45 nights between 25 Dec. 1998 and 22
Feb. 1999 with the Wide Field Imager attached to the 2.2m MPG/ESO
telescope at La Silla in Chile.  Details of the data acquisition, analysis
and additional results will be reported elsewhere \citep{her01}. 
Here we note that the field surveyed was a $33\arcmin$ x $34\arcmin$
rectangle centered approximately on $\theta^1$ Ori C, making it nearly
coincident with the ONC defined by Hillenbrand and Hartmann (1998). 
Photometry was obtained on 2,294 stars extending to I$\sim$18.  A search
for periodicity was carried out using the Lomb-Scargle technique and
standard assessments of false alarm probabilities  
\citep{her00,reb01}.  Of the 404 periodic stars identified in our sample,
111 had been previously discovered by \citet{her00} 
or \citet{sta99} and 99 of these were found to have identical periods to within
the errors of the determinations ($\sim$3\%).  A list of periodic 
stars is available upon request to the first author. Eleven of the 12 stars
which had period disagreements were cases of either harmonics or beat
periods masquerading as fundamentals.  The quality and quantity of the data
obtained for this study is unprecedented and that is responsible for our
success in nearly tripling the number of rotation periods now known for
stars in the ONC. In particular, our study extends to fainter stars and,
therefore, lower masses than have been probed heretofore and it is that
aspect of the results which we primarily discuss in this {\it Letter}.  

Fig. 1
shows the distribution of rotation periods for 335 periodic stars in our
sample which have masses determined by \citet{hil97}.  Her
determinations are based on a comparison of each star's location in the HR
diagram with PMS models of \citet{dm94}.  Masses are, of
course, model-dependent and these could be systematically in error by
perhaps as much as 50\%.  However, all models of PMS stars indicate that
lower effective temperatures correspond to lower
masses, so the sense of the variation of rotation with mass in Fig.  1 is
model independent.  It is evident in the figure that the range of rotation
rates is very large, regardless of the mass range considered.  While the
extreme values on both the high and low period ends may be questionable
(due to possible effects of aliasing and harmonics) it is fair to say that
rotation periods in this cluster span at least a range of 0.8 to 12 
days, independent of mass.  What may be less obvious at first
inspection, on account of that wide range, is that there is a definite
change in the nature of the rotation period distribution with mass.  Part
of that trend is indicated by the heavy solid line which shows the median
period within a sliding sample defined by a mass range $\pm$0.05 M$_\odot$
at 8 central values from 0.1 to 0.4 M$_\odot$.  This stastic and 
sampling interval
were chosen for their robustness in the range M $<$ 0.4 M$_\odot$, the focus
of this paper.  

To make the trends clearer and assess their statistical
significance we show, in Figure 2, histograms of the rotation period
distribution in three mass ranges.  A double-sided Kolmogorov-Smirnof
test indicates that the probability that the high mass sample was drawn
from the same parent population as the low mass sample is less than $3
\times 10^{-9}$. This result is independent of the binning chosen to 
display the histograms. Clearly, the higher mass
ranges show a bimodal distribution, as first discovered by \citet{ah92} 
and later confirmed by \citet{ch96}.  Bimodality can
be crudely quantified using the Double Root Residual test  
\citep{gb91}
which, in this case, indicates for the highest mass range that the distribution differs
from uniformity at greater than the 3$\sigma$ level.  Clearly, the middle
sample shows a mixture of attributes. 
We conclude that the rotation period distribution in the ONC is bimodal for
stars with M $\geq$ 0.25 M$_\odot$ and unimodal for lower mass stars,
confirming the results of \citet{her00}.  The interesting new result is that 
the majority of stars with M$<$0.3 M$_\odot$ clearly rotate faster than the 
majority of higher mass stars and we turn now to a discussion of this issue.

\section{Analysis and Discussion}

In an attempt to better understand these results, we first translate the
observed quantity, rotation period (P), into the fundamental physical
quantity, specific angular momentum (j).  Assuming that convection 
enforces rigid rotation in low mass stars, even at 1 My,
the specific angular momentum of such a star with mass M,
and radius R, is 
$$ j = {J \over M} = {{I\omega} \over M} = {{2 \pi k^2 R^2 } \over P}$$ 
where J is the total angular momentum, I is the moment of inertia, $\omega$ is the angular velocity and 
k is the radius of gyration in units of the stellar radius.  We assume
homologous stars and adopt k=0.44, appropriate to the 1 My old
PMS models of \citet{kri97}.  It follows that $$j = {
{1.2 \times 10^{17} (R/R_\odot)^2} \over {(P /day)}}\ 
cm^2\ s^{-1}.$$ A contracting PMS star will spin up roughly as P $\propto$
R$^2$ if it conserves angular momentum.  At no time, however, can a star
spin more rapidly than a balance between gravitational and centrifugal
forces at its surface will allow.  This criterion defines a critical period
(P$_{crit}$) which applies to the axial rotation of a rigid sphere, namely
$$P_{crit} = {0.1156 \times {(R/R_\odot)^{3/2} \over (M 
/M_\odot)^{1/2}}}\ d.$$ which is $\sim$0.5 d for a 2 R$_\odot$, 0.5 
M$_\odot$ 
 PMS star.  A
corresponding critical specific angular momentum (j$_{crit}$) 
may be defined by inserting
P$_{crit}$ into the expression for j above.  

In Figure 3, we show the observed value of specific angular momentum
(j$_{obs}$) as well as j$_{crit}$ for 
stars in our ONC sample. In calculating j$_{obs}$ for M$<$0.4 M$_{\odot}$ we
have used the median period, shown in Figure 1, and the median radius (from
\citet{hil97}) for the same moving samples.
Dashed lines in this range show the locations of the quartiles of the 
sample (see also Figure 1). For the higher mass stars, we show
two values of j$_{obs}$ corresponding to the two peaks in the  
period distribution and we have adopted the model radii of 
\citet{dm94} 
because the sample is too sparse to justify calculation of a moving 
median. It is clear from this figure that the specific angular 
momentum of a ``typical'' PMS star shows very little dependence on 
mass over the one decade interval 0.1-1.0 M$_{\odot}$. Here we define 
``typical'' as a star having the median rotation rate at 
lower mass or having P$\sim$8 d at higher mass. (As Fig. 1 shows, about 2/3 of 
the higher mass stars in the ONC fall in the slower rotating peak of 
the period distribution.) 
There is evidence for a small (factor of $\sim2$) 
increase in j with decreasing mass in the range 0.1$<$M$<$0.4 
M$_\odot$, although a constant value is only barely inconsistent with 
the errors.  

Three estimates of j$_{crit}$ are also shown in Fig. 3 for comparison 
with the observations. It may be seen from the expressions above that  
j$_{crit} \propto (M/R)^{1/2}$, so it requires masses but is not 
highly sensitive to them. 
The values represented by the solid line correspond precisely to the 
observations. That is, for M$<$0.4 M$_{\odot}$, they are based on the 
same moving median samples, with radii and masses taken from \citet{hil97} 
and for the higher mass stars they are based entirely on the 1 My old models of 
\citet{dm94}. The model results are shown 
extended into the low mass range by the dotted line in Fig. 3, 
revealing only a small difference, as expected. 
To check whether a completely different set of models would give 
similar results, we also show, as a dashed-dotted line,
calculations of j$_{crit}$ based on the 
1 My old PMS models of \citet{ps99}. Clearly the results are in good 
agreement and indicate a significant trend of decreasing j$_{crit}$ 
with decreasing mass. 

It is clear from Fig. 3 that lower mass
stars in the ONC rotate at closer to their critical rates than do their 
higher mass 
counterparts. Specific angular momentum exceeds 25\% of the critical value 
for the lowest
masses (M$\sim$0.1 M$_{\odot}$) in our sample, compared to $\sim$5\%, 
which is typical of solar-like stars.  This
result is model-dependent in the sense that the j$_{obs}$ curve would 
translate along
the mass axis if different models were employed to infer masses from 
luminosity and effective temperature. 
However, the general shape of the curve (i.e. fairly flat)
is independent of the PMS model chosen for this transformation. The 
observed value of j$_{obs}$ 
depends only on P and R, and R is determined from luminosity and 
effective temperature without reference to stellar models.  
(If there were a mass-dependent systematic error in R, this would 
affect the shape of j$_{obs}$ but that possibility is not given 
serious consideration here.) We conclude, therefore, that lower
mass stars in the ONC rotate at rates much closer to their critical angular
velocity than higher mass stars, and they do so for two reasons. First,
j increases somehwat at lower masses, or at the very least remains 
constant, and second, j$_{crit}$ decreases with mass, which is mainly a result
of the decrease in radius predicted by the models.  

Our observations in the ONC are
interesting from two perspectives.  First, they can serve as the initial
conditions for theoretical models of angular momentum evolution which hope
to explain the 30-600 My old cluster data.  The calculation of such models
is beyond the scope of this paper.  Here we note, however, that the basic
features of the angular velocity distribution with mass in the Pleiades
recently discussed by \citet{ter00} are rather precisely mimicked
in our data on the ONC. In particular, both clusters exhibit a bimodal
period distribution for more massive stars and a unimodal distribution with
increasingly rapid rotation at lower masses.  The mass at which the nature of
the distribution changes is, possibly, higher in the Pleiades but 
could be the same, given the large uncertainties in both clusters.  
In any event, these data suggest that
future studies of rotational evolution of low mass stars such as that of
\citet{sil00} should consider a mass-dependent
starting angular velocity.  

The second issue raised by our data is this: can we understand the ONC
rotation distribution in terms of the mechanism(s) which establish the
``initial" angular momentum of a star during the first $\sim$1 My of its
life?  Bodenheimer (1995) has reviewed those mechanisms and it is clear
that the mass dependence, or lack of a strong one, reported here could be 
an important clue to
sorting out the relevant processes.  It has long been thought that magnetic
fields must play a key role in removing angular momentum from proto-stars
and PMS stars.  This view is certainly supported by the fact that we detect
the spot variations of PMS stars; they must have large, stable cool spots
on their surfaces and this is {\it prima facie} evidence that they have
strong, well organized magnetic fields.  The disk-locking mechanism 
of \citet{cam90},
\citet{kon91} and \citet{shu94} provides a plausible explanation for
the fact that $\omega$ $\approx$ 0.05$\omega_{crit}$ for a typical T
Tauri star of 0.5 M$_\odot$ \citep{os95}.  It also accounts
nicely for the bimodal distribution of the more massive ONC stars 
\citep{ch96,her00} and for the wide range in angular momenta of
cluster stars \citep{bar01}.  Could it also explain the fact that lower
mass stars in the ONC (and the Pleiades) rotate more rapidly, but 
with only slightly enhanced or, perhaps, a constant value of j?   
It is not
obvious to us that there is anything fundamental in the disk-locking theory
which predicts a weakly mass-dependent rotation law of the form we 
observe, but suspect that it could easily accomodate one.  
We leave these considerations to 
future investigations, with the hope that the observational guidance 
provided here may prove useful in understanding the complex issues of 
angular momentum transport during the PMS phase.



\acknowledgments
We thank S. Barnes for helpful discussions on the topic of this paper.  It
is a pleasure to thank K. Meisenheimer, R. Wackermann and Ch. Wolf for assistance
with the observations.  This work was partially supported by a grant from
the NASA Origins program.




\clearpage
\begin{figure}
    \figurenum{1}
    \epsscale{0.8}
    \plotone{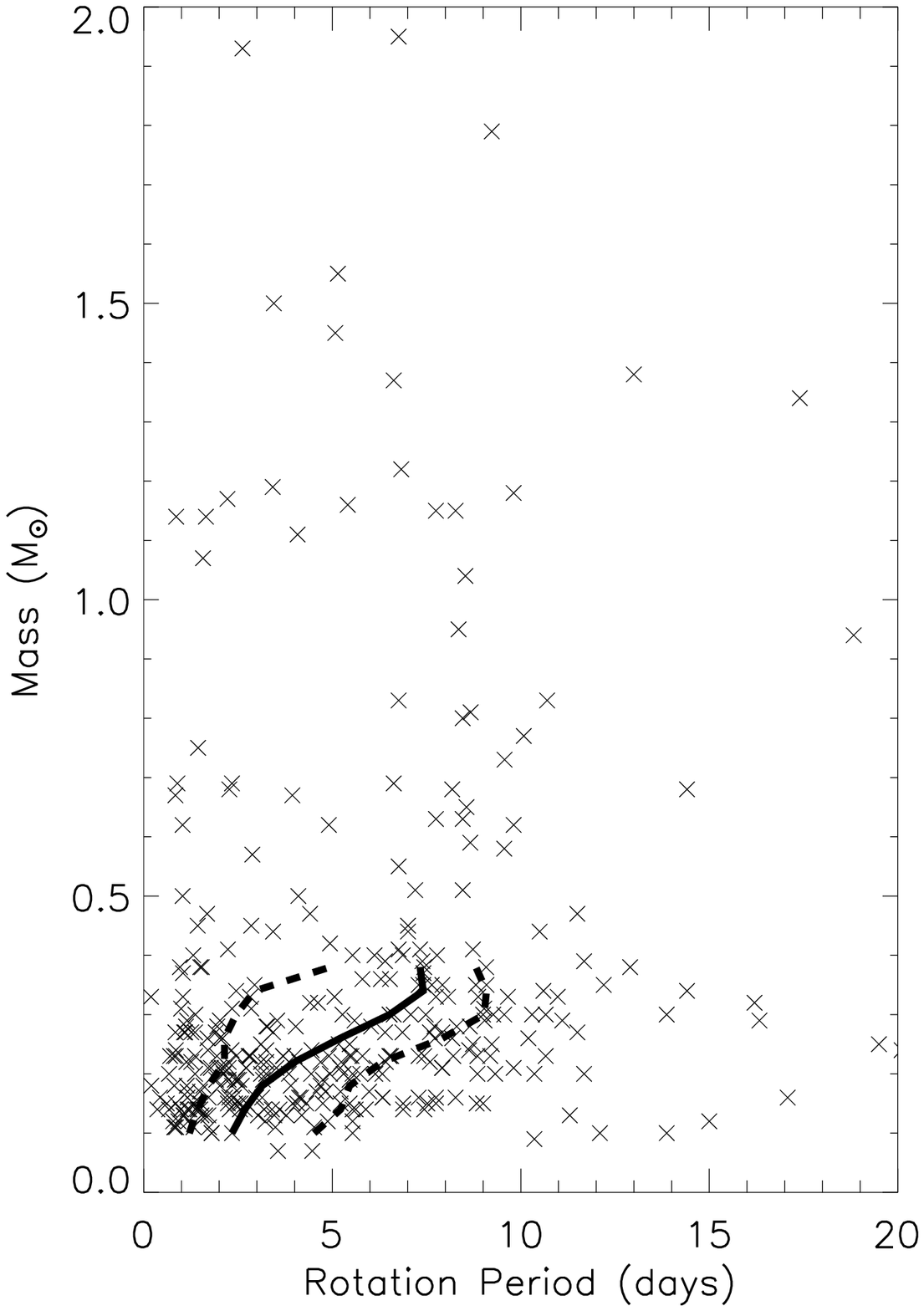}
    \caption{Rotation period versus mass for stars in the ONC. The heavy solid
    line is the moving median defined in the text and the 
    dashed lines are the corresponding quartiles.  
    The moving median is not shown for M $<$ 0.4 M$_\odot$ because the 
    relatively small number of stars makes it an unreliable statistic 
    in that range.}
    \end{figure}
\clearpage

\begin{figure}
    \figurenum{2}
    \epsscale{0.6}
    \plotone{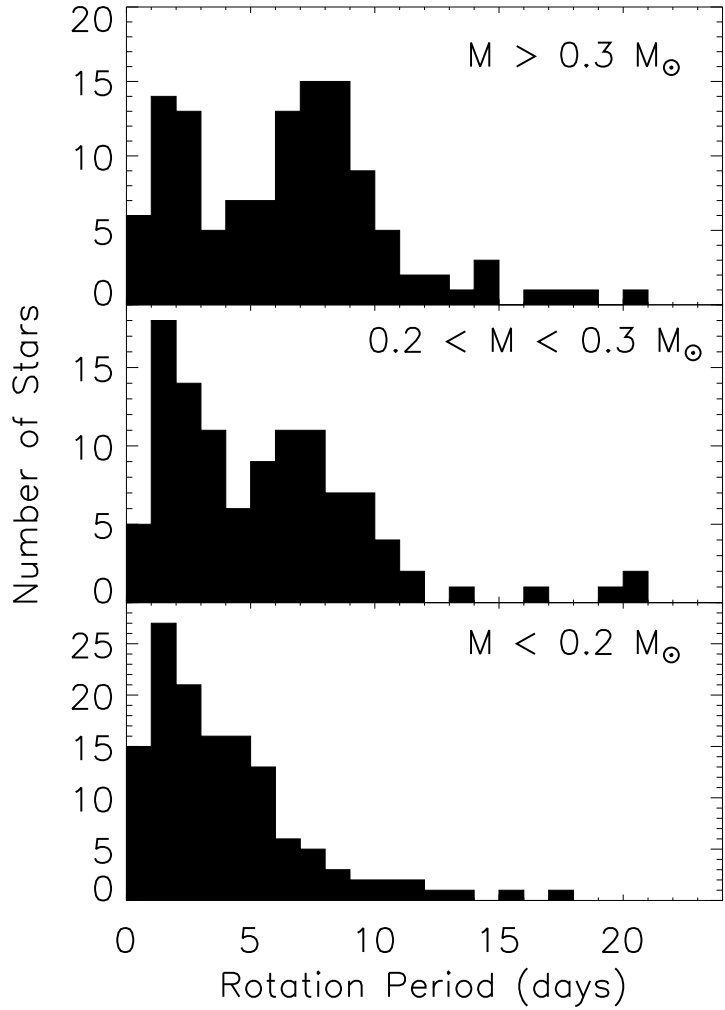}
    \caption{Rotation period distributions for stars in three mass ranges
    chosen to illustrate the data trends seen in Figure 1.  A bimodal
    distribution is clearly seen in the highest mass range and a unimodal
    distribution consisting of rapid rotators is seen in the lowest mass range. 
    The intermediate range shows a mixture of characteristics.}
    \end{figure}
\clearpage    
 
\begin{figure}
    \figurenum{3}
    \epsscale{0.8}
    \plotone{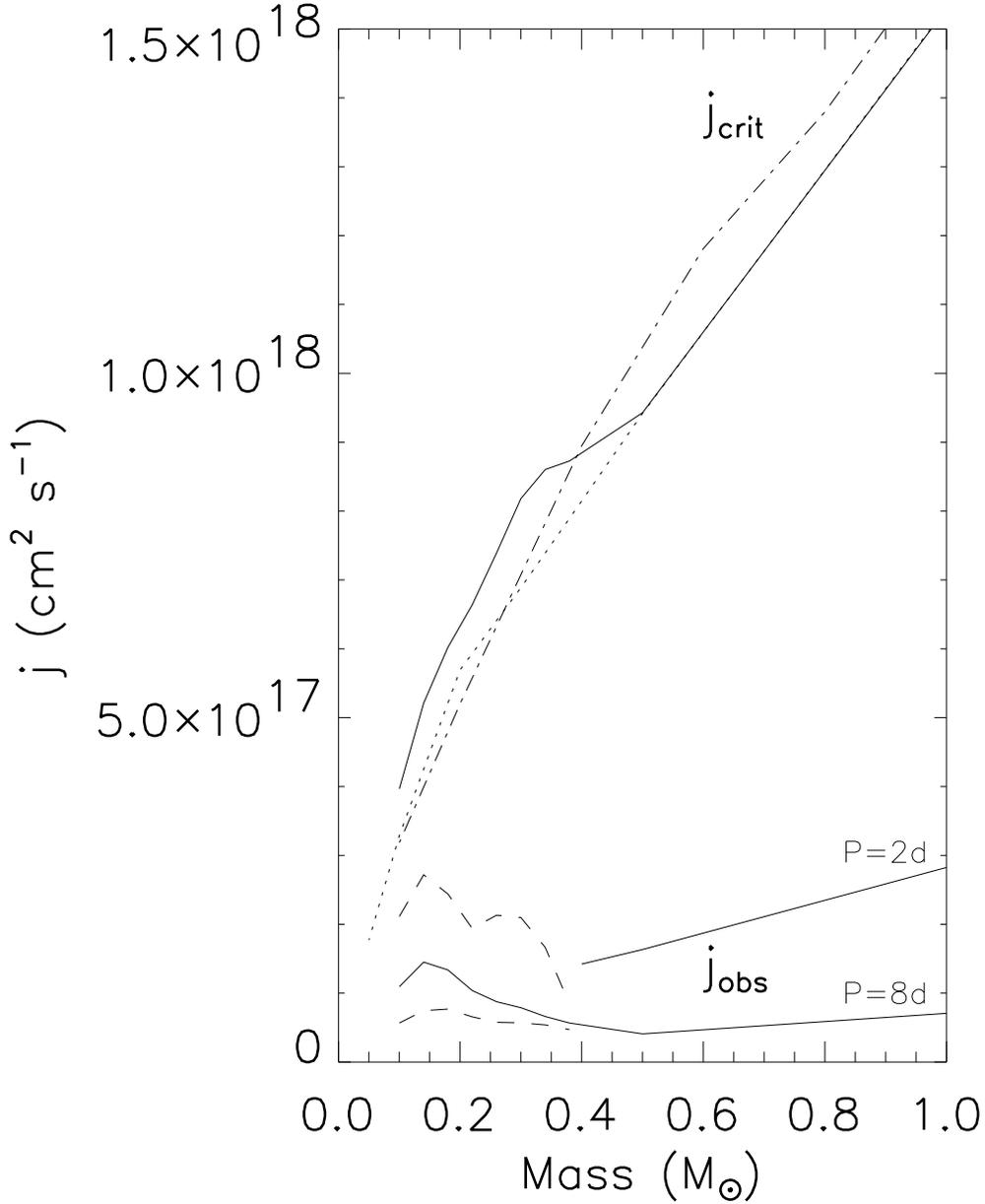}
    \caption{The specific angular momentum (j=J/M) as a function of 
    mass. The lower solid line shows the observations, including 
    their bimodal character with periods of 2 and 8 days for M$>$0.4 
    M$_{\odot}$. The dashed 
    lines are an estimate of the uncertainty based on the quartile 
    values (see Fig. 1).  The upper solid 
    line shows the value of j$_{crit}$ for a rigid sphere with the mass and 
    radius (calculated as a moving median over the same sample 
    described in the text) rotating at its critical velocity. The 
    dotted and dashed-dotted lines show j$_{crit}$ for model PMS stars from 
    \citet{dm94} and \citet{ps99}, respectively, rotating at 
    critical velocity. It is clear from this figure that lower mass stars 
    rotate at a
    greater percentage of their critical angular velocity than do higher mass
    stars in the ONC.}
    \end{figure}
\clearpage






\end{document}